\documentstyle[12pt,epsfig]{article}

\newcommand{\be}{\begin{equation}}
\newcommand{\ee}{\end{equation}}

\newcommand{\dlt}{\delta}

\newcommand{\br}{{\bf r}}
\newcommand{\bS}{{\bf S}}

\newcommand{\bB}{{\bf B}}

\newcommand{\bn}{{\bf n}}
\newcommand{\bt}{\beta}

\newcommand{\al}{\alpha}

\newcommand{\gm}{\gamma}
\newcommand{\om}{\omega}

\newcommand{\Gm}{\Gamma}

\begin{document}

\begin{center}
{\Large{\bf Coherent spin relaxation in molecular
magnets} \\ [5mm]
V.I. Yukalov$^1$, V.K. Henner$^{2,3}$, and P.V. Kharebov$^2$} \\ [3mm]

{\it
$^1$Bogolubov Laboratory of Theoretical Physics, \\
Joint Institute for Nuclear Research, Dubna 141980, Russia\\ [3mm]
$^2$Perm State University, Perm 614000, Russia \\ [3mm]
$^3$University of Louisville, Louisville 40292, KY, USA
}

\end{center}

\vskip 2cm

\begin{abstract}

Numerical modelling of coherent spin relaxation in nanomagnets, formed by
magnetic molecules of high spins, is accomplished. Such a coherent spin
dynamics can be realized in the presence of a resonant electric circuit
coupled to the magnet. Computer simulations for a system of a large number
of interacting spins is an efficient tool for studying the microscopic
properties of such systems. Coherent spin relaxation is an ultrafast
process, with the relaxation time  that can be an order shorter than the
transverse spin dephasing time. The influence of different system
parameters on the relaxation process is analysed. The role of the sample
geometry on the spin relaxation is investigated.

\end{abstract}

\vskip 1cm

{\bf PACS}: 76.20.+q, 75.30.Gw, 75.45.+j, 75.50.Xx, 76.90.+d

\newpage

\section{Introduction}

A polarized magnet placed in an external magnetic field, whose
direction is opposite to the sample magnetization, comprises a
strongly nonequilibrium system. The relaxation of spins to their
equilibrium position can occur in different ways. The simplest
process is the slow relaxation during the time $T_1$ caused by
the spin-lattice coupling [1], when the total magnetization
tends to zero. This is an incoherent process,
since $T_1$ is usually much longer than the spin dephasing time
$T_2$. A straightforward way to make the spin relaxation coherent
would be by imposing a strong transverse field simultaneously
driving all spins, which would represent the process of {\it
free induction} [1], lasting for the time $T_2$. This, however,
is a rather trivial process, when spins are practically
independent of each other.

A more elaborate situation arises if the magnet is inserted into
a magnetic coil of a resonant electric circuit. Then the magnetic
field, induced by the coil, provides an efficient feedback mechanism
organizing the coherent motion of spins [2]. It is possible to realize
six different regimes of coherent spin relaxation in a polarized
sample coupled to a resonant circuit: {\it Collective induction,
maser generation, pure superradiance, triggered superradiance, pulsing
superradiance}, and {\it punctuated superradiance}. A detailed
description of these regimes is given in the review articles [3--5].

Collective induction and maser generation with spins were realized
in a number of experiments [6--9]. Pure spin superradiance was,
first, observed in Dubna [10,11] and later confirmed by the groups
in St. Petersburg [12,13] and Bonn [14]. Pulsing superradiance was
demonstrated by experiments in Z\"urich [15,16]. The regime of
punctuated superradiance was suggested in Ref. [17] and, to our
knowledge, has not yet been realized experimentally. A comprehensive
survey of experiments can be found in Refs. [4,5,18].

It is necessary to emphasize the principal role of the resonant
electric circuit, coupled to the spin sample, for realizing the
regimes of spin superradiance. This makes the fundamental difference
between the spin superradiance and the atomic superradiance [19].
The latter can be achieved in a resonatorless system [19--21],
though a resonant cavity can enhance the effect [22,23]. Contrary
to this, the spin superradiance, occurring in the radiofrequency
range, cannot be realized without a resonator, which is due to the
destructive role of the dipole spin interactions and to the absence
of the feedback mechanism collectivizing the spin motion. This basic
difference was emphasized in Ref. [19] and thoroughly explained
in Refs. [5,24,25]. One should also distinguish between coherent
transient effects, caused by intense alternating external fields,
and having much in common for optical atomic systems [20,21],
gamma radiation [26--29], as well as for spin samples [30--32],
and superradiant phenomena, when the {\it self-organization} of
coherence is the basic origin of the arising superradiance
[25,33,34].

Here we concentrate our attention on the superradiant regime of spin
motion, which requires the presence of the resonant electric circuit,
providing the feedback mechanism for the collective self-organization
of spin motion. The regimes of free induction, collective induction,
and triggered spin superradiance could be considered in the frame of
the phenomenological Bloch equations supplemented by the Kirchhoff
equation for the circuit [35--38]. These phenomenological equations,
however, are not applicable for describing pure spin superradiance,
when the coherent motion of spins develops in a self-organized way
from initially chaotic spin fluctuations. The complete theory, based
on the microscopic spin Hamiltonian, and describing all regimes of
spin relaxation has been developed in Refs. [18,25,33,34,39] and
expounded in detail in the review articles [4,5]. This theory is in
good agreement with experiment [4,5] as well as with numerical computer
simulations [40--42] accomplished for nuclear or electron spins one half,
$S=1/2$.

There also exists a wide class of paramagnetic materials formed by
magnetic molecules [43], whose effective ground-state spins can reach
rather high values of $S\sim 10$. These molecules compose molecular
magnets whose properties are described at length in Refs. [5,18,44,45].
For example, the molecules Mn$_{12}$ and Fe$_8$ possess the spin $S=10$.
The magnetic cluster compound Mn$_6$O$_4$Br$_4$(Et$_2$dbm)$_6$ has
the total spin $S=12$ [46]. And the magnetic molecule
Cr(CNMnL)$_6$(CIO$_4$)$_9$, where L stands for a neutral pentadentate
ligand, displays the effective spin $S=27/2$ [45].

Molecular magnets, formed by high-spin magnetic molecules, exhibit
a rather strong magnetic anisotropy and a gigantic relaxation time
of their magnetization, which reaches about two months in zero
magnetic field at low temperature [5,18,44,45]. The magnetization
relaxation is mainly due to the phonon-assisted mechanism caused by
spin-phonon interactions [47,48]. The fast superradiant-type relaxation
of magnetization is theoretically possible for molecular magnets,
provided they are placed in a sufficiently strong external magnetic field
and necessarily coupled to a resonant circuit [5,18,25,49].

The main points, distinguishing the relaxation in magnetic molecules,
studied earlier [44,45,50--57], from the process to be investigated in the
present paper, are as follows.

First of all, the usually studied spin relaxation in magnetic molecules
is the process of quantum tunneling occurring in individual molecules
[50--57]. Contrary to this, we shall be interested in coherent, principally
collective effects, due to strong correlations between many molecules,
which is caused by the resonator feedback field.

Quantum tunneling in separate molecules is known to be phonon assisted,
with an essential phonon influence on the relaxation process [47,48,58].
But in the case of spins strongly correlated through the resonant feedback
field, spin-phonon interactions are not as important, the basic dynamics
being governed by the resonator feedback field.

The single-molecule spin relaxation is not connected with a fixed
particular frequency [59], because of a sufficiently strong magnetic
anisotropy and a varying external magnetic field [5,19,24,25]. While
to realize the collective coherent spin relaxation necessary requires
the situation close to resonance between the Zeeman frequency and the
resonator natural frequency.

Magnetic avalanches are usually accomplished by sweeping the longitudinal
magnetic field with a rather slow sweeping rate. A typical field sweeping
rate used in experiments [50--55] is about 0.1 T$/$s. In our case, we
consider a fixed external magnetic field, with a well defined Zeeman
frequency.

The avalanche time of the magnetic moment in magnetic molecules is
between $10^{-3}$ s and $1$ s [50--57,59]. This is rather slow process,
as compared to the coherent spin relaxation, which is an ultrafast
process, with characteristic relaxation times about $10^{-10}-10^{-13}$
s [5,18,24,25,49].

The peculiarities of the fast superradiant-type spin relaxation in
molecular magnets can be successfully analyzed by means of computer
modelling, which, to our knowledge, has not yet been accomplished.
This approach provides a very efficient tool for studying the
microscopic properties of spin system. And it is the aim of the present
paper to describe the results of computer simulations for analyzing the
features of the fast coherent relaxation of molecular spins, typical of
the high-spin molecular magnets.

The outline of the paper is as follows. In Sec. II, we present the
main definitions and equations to be employed in our computer modelling.
In Sec. III, the results for a bulk sample are analyzed. Since the
dipolar spin interactions are anisotropic, it is interesting to study
the related anisotropic geometric effects for different shapes of
the magnetic samples. These geometric effects are investigated for
a chain of molecular spins oriented either along the external
magnetic field or perpendicular to it and for spin planes, with the
external magnetic field being either perpendicular to it or lying in
that plane. Section IV contains conclusions.

\section{Basic Definitions and Equations}

We consider a spin sample characterized by the molecular spin vectors
$\bS_j=\{ S_j^x,S_j^y,S_j^z\}$ associated with the lattice sites
enumerated by the index $j=1,2,\ldots,N$. An external magnetic field
is directed along the $z$-axis,
\be
\label{1}
\bB_0 = B_0\;{\bf e}_z \; .
\ee
This defines the Zeeman frequency
\be
\label{2}
\om_0 \equiv -\; \frac{\mu_0}{\hbar} \; B_0 = \frac{2}{\hbar}\;
\mu_B \; B_0 \; ,
\ee
in which $\mu_0=-2\mu_B$ is the electron magnetic moment, with $\mu_B$
being the Bohr magneton. In general, similarly to the magnetic-resonance
setup, there can exist the transverse magnetic field, directed along the
$x$-axis,
\be
\label{3}
\bB_1 = B_1 {\bf e}_x \; , \qquad B_1 = h_0 + h_1 \cos \om t \; ,
\ee
and consisting of a constant field $h_0$ and an alternating field
$h_1\cos\om t$.

Molecular magnets possess the single-site magnetic anisotropy
characterized by the anisotropy parameter $D$, which defines the
anisotropy frequency
\be
\label{4}
\om_D \equiv \left ( 2S - 1 \right ) \; \frac{D}{\hbar} \; ,
\ee
where $S$ is the molecular spin. The magnetic anisotropy exists for high
spins, playing an important role, while for $S=1/2$ it disappears,
according to Eq. (4).

Spins interact with each other through the dipolar forces characterized
by the dipolar tensor
\be
\label{5}
D_{ij}^{\al\bt} \equiv \frac{\mu_0^2}{r_{ij}^3} \; \left (
\dlt_{\al\bt} - 3 n_{ij}^\al \; n_{ij}^\bt \right ) \; ,
\ee
in which
$$
r_{ij} \equiv |\br_{ij} | \; , \qquad \bn_{ij} \equiv
\frac{\br_{ij}}{r_{ij}} \; , \qquad \br_{ij} \equiv \br_i -\br_j \; .
$$
For what follows, it is convenient to introduce the dipolar coefficients
\be
\label{6}
a_{ij} \equiv D_{ij}^{zz} \; , \qquad b_{ij} \equiv \frac{1}{4} \left (
D_{ij}^{xx} - D_{ij}^{yy} - 2i D_{ij}^{xy} \right ) \qquad
c_{ij} \equiv \frac{1}{2}\left ( D_{ij}^{xx} - i D_{ij}^{yz} \right ) \; ,
\ee
having dimensions of energy.

The spin sample is inserted into a magnetic coil of an electric circuit
characterized by the circuit damping $\gm$ and the circuit natural
frequency $\om$. Moving spins generate electric current in the coil,
which, in turn, produces the feedback magnetic field ${\bf H}$ acting
on the spins. The generated electric current is described by the
Kirchhoff equation. Choosing the coil axis along the $x$-axis, so that
${\bf H}=H{\bf e}_x$, the Kirchhoff equation can be rewritten [33,34]
as the equation for the feedback magnetic field $H$,
\be
\label{7}
\frac{dH}{dt} + 2\gm H + \om^2 \; \int_0^t \; H(t')\; dt' =
-4 \pi\eta \; \frac{dm_x}{dt} \; ,
\ee
where the effective electromotive force in the right-hand side of Eq.
(7) is produced by moving spins, with the average magnetization
\be
\label{8}
m_x = \frac{\mu_0}{V}\; \sum_{j=1}^N < S_j^x > \; ,
\ee
$V$ being the sample volume. The filling factor $\eta$ in the right-hand
side of Eq. (7) is approximately equal to $\eta=V/V_c$, where $V_c$ is
the coil volume. For what follows, without the loss of generality, we may
assume the dense filling, with $\eta=1$. Instead of Eq. (7), we can use
the equivalent differential equation
\be
\label{9}
\frac{d^2H}{dt^2} + 2 \gm \; \frac{dH}{dt} + \om^2 H =
- 4\pi\; \frac{d^2 m_x}{dt^2} \; ,
\ee
in which we set $\eta=1$.

All possible attenuation mechanisms have been carefully described in
Ref. [25]. Those that influence the spin motion are as follows. The
longitudinal attenuation
\be
\label{10}
\Gm_1 = \gm_1 +\gm_1^*
\ee
is the sum of the spin-lattice attenuation $\gm_1$, caused by
spin-phonon interactions, and of the polarization pump rate $\gm_1^*$,
due to a stationary nonresonant pump, if any. The total transverse
attenuation is
\be
\label{11}
\Gm_2 = \gm_2 \left ( 1 - s^2 \right ) + \gm_2^* \; .
\ee
This includes the homogeneous broadening $\gm_2$, renormalized by the
factor $(1-s^2)$, appearing in the case of strongly polarized spin systems
[1,25], with $s$ being the average spin polarization reduced to the number
of spins $N$ and to the spin value $S$. The last term $\gm_2^*$ is the
static inhomogeneous broadening.

In order to represent the equations of spin motion in a compact form,
it is convenient to introduce the ladder spin components
\be
\label{12}
S_j^- \equiv S_j^x - i S_j^y \; , \qquad
S_j^+ \equiv S_j^x + i S_j^y \; .
\ee
Also, we shall use the following notation:
$$
\xi_i^0 \equiv \frac{1}{\hbar} \; \sum_{j(\neq i)} \left (
a_{ij} S_j^z + c_{ij}^* S_j^- + c_{ij} S_j^+ \right ) \; ,
$$
\be
\label{13}
\xi_i \equiv \frac{1}{\hbar} \; \sum_{j(\neq i)} \left (
2 c_{ij} S_j^z \; - \; \frac{1}{2}\; a_{ij} S_j^- +
2b_{ij} S_j^+ \right ) \; .
\ee
The effective force, acting on the $j$-spin, can be written as
\be
\label{14}
f_j \equiv -\; \frac{i}{\hbar} \; \mu_0 (B_1 + H)\; + \;
\xi_j \; .
\ee

The derivation of equations of motion for the spin variables $S_j^-$,
$S_j^+$, and $S_j^z$ has been described in great detail in Refs.
[4,5,25]. The resulting equation for $S_j^-$ reads as
\be
\label{15}
\frac{dS_j^-}{dt} = - i \left ( \om_0 + \xi_j^0 - i\Gm_2
\right ) S_j^- \; + \; f_j S_j^z \; + \; i\; \frac{\om_D}{S}\;
S_j^z S_j^- \; .
\ee
The equation for $S_j^+$ is conjugate to Eq. (15). And the equation
for $S_j^z$ is
\be
\label{16}
\frac{dS_j^z}{dt} = -\; \frac{1}{2} \left ( f_j^+ S_j^ - \; +
\; S_j^+ f_j \right ) - \Gm_1 \left ( S_j^z - \zeta \right ) \; ,
\ee
where $\zeta$ is the stationary spin polarization. From Eqs. (15)
and (16), with notation (12), one can always return to the evolution
equations for $S_j^x$, $S_j^y$, and $S_j^z$.

In numerical simulations, one treats the spins $\bS_j$ as classical
vectors [40--42]. It is convenient to work with the reduced quantities
characterizing the reduced transverse spin variable
\be
\label{17}
u \equiv \frac{1}{SN} \; \sum_{j=1}^N \; S_j^-
\ee
and the reduced longitudinal spin variable
\be
\label{18}
s \equiv \frac{1}{SN} \; \sum_{j=1}^N \; S_j^z \; .
\ee
The spin variables (17) and (18) characterize the collective properties
of a large number of magnetic molecules composing the molecular magnet.
The time evolution of these variables is prescribed by the equations of
motion (15) and (16). This picture of collective spin motion is a
generalization of the evolution equations for a single magnetic molecule.
The study of collective coherent effects is the main aim of the
present paper.

In our numerical simulations, we solve the spin evolution equations (15)
and (16) for a finite number of spins $N$. The resonator feedback field
is given by Eq. (9), with the initial conditions
\be
\label{19}
H(0) = 0 \; , \qquad \dot{H}(0) = 0 \; ,
\ee
where the overdot implies the time derivative of $H$. The spin
variables $S_j^\al$ at the initial time are distributed randomly over
the sample, so that to obtain a prescribed value $s(0)$ of the spin
polarization (18), while variable (17), for sufficiently high initial spin
polarization, being negligible,
\be
\label{20}
s(0) =s_0 \; , \qquad u(0) = 0 \; .
\ee

The external magnetic field (1), with $B_0>0$, is aligned with
${\bf e}_z$. For the initial spin polarization $s_0>0$, the magnetic
moment of the molecular sample
$$
{\bf M}(0) = NS \mu_0 s_0 {\bf e}_z = -2 NS \mu_B s_0{\bf e}_z
$$
is opposed to ${\bf e}_z$. That is, the considered molecular magnet
is prepared in a strongly nonequilibrium initial state, from which it
relaxes to a stationary state.

\section{Results of Computer Simulation}

First, we consider the coherent spin relaxation in bulk samples, where
spins are located in lattice sites of a cubic lattice. Computer simulations
are accomplished for $N=125$ spins. For larger values of $N$, the results
are qualitatively similar. The periodic boundary conditions have been
imposed. Since our aim is to study the self-organized process, we set zero
the transverse field $B_1=0$. We assume that the spin sample is without
defects, so that the inhomogeneous broadening is negligible, $\gm_2^*=0$.
For low temperatures, the spin-lattice interaction in molecular magnets
is very weak, with the longitudinal relaxation time $T_1$ reaching months
(see review articles [5,18,44,45]), because of which the attenuation
parameter $\Gm_1$ plays no role, and we can set $\Gm_1=0$. It is
convenient to deal with dimensionless quantities measuring all
frequencies in units of $\gm_2$, so that we set $\gm_2=1$. And we
shall measure time in units of $\gm_2^{-1}$, that is, in units of $T_2$.

First and foremost, we have to stress the necessity of the resonant
circuit. When the latter is absent, there is no fast relaxation at all.
Then there could exist only the very slow polarization decay during the
time $T_2$, which is caused by spin interactions.
The same slow relaxation happens if there is a circuit, but there is no
resonance between its natural frequency $\om$ and the Zeeman frequency
$\om_0$. Therefore, in what follows, we always set $\om=\om_0$. This
resonance condition is necessary, though not sufficient. To realize an
effective spin reversal, it is important that $\om_0$ be much larger than
the anisotropy frequency (4). The condition $\om_0\gg\om_D$ ensures that
the anisotropy does not induce an effective detuning from the resonance
[5,18,24,25].

Figure 1 illustrates how the spin reversal depends on the value of the
Zeeman frequency $\om_0$. The larger the latter, the more pronounced is
the spin reversal.

The resonator damping $\gm$ defines the resonator ringing time
$\tau\equiv1/\gm$, during which the magnetic sample effectively interacts
with the resonator. The relation between the resonator ringing time $\tau$
and the transverse relaxation time $T_2$ essentially influences the spin
reversal. When $\tau=T_2$, then there is a permanent exchange of energy
between the spin sample and the resonator, so that the spin polarization
oscillates around zero. When $\tau=0.1$ $T_2$, there occurs a well
pronounced reversal, hence the value $\gm=10$ is optimal for the latter.
And if $\tau=0.02$ $T_2$, then the effective interaction time between the
sample and resonator is too short to realize a good reversal of
polarization. The corresponding three qualitatively different cases ate
shown in Fig. 2.

The magnitude of spin reversal also depends on the initial polarization.
The larger the initial value $s_0$, the stronger the spin reversal, which
is illustrated in Fig. 3.

Magnetic anisotropy is an obstacle for the coherent spin relaxation.
The larger the value of $\om_D$, the smaller the spin reversal, as is
demonstrated in Fig. 4.

Dipole spin interactions is also a factor suppressing spin coherence. This
is illustrated by Fig. 5, where the behavior of spin polarization for the
case with dipole interactions is compared with that one, for which the
dipole interactions are switched off by setting $D_{ij}^{\al\bt}=0$.

It is interesting that switching off the dipole interactions yields the
figures that are very close to those obtained by
the reduction of spin from $S=10$ to $S=1/2$. Thus the dashed line in
Fig. 5, where $\gm=10$, for $S=10$ can be repeated not by setting the
dipole tensor to zero but by reducing the spin to $S=1/2$. In Fig. 6, we
show the behavior of spin polarization for $S=10$ and $S=1/2$ for $\gm=1$.
Again, switching off the dipole interactions for $S=10$ yields the dashed
curve corresponding to $S=1/2$ with dipole interactions.

Dipole interactions are anisotropic. Therefore, their influence on the
relaxation process can be different for the samples of different shapes
and of different orientations with respect to the external magnetic field
and with respect to the resonator feedback field. To analyze these
differences, we study the spin relaxation, under the same system
parameters, but for different samples. We consider the chain of spins
oriented either along the external magnetic field, i.e., along the
$z$-axis, or along the feedback field, that is, along the $x$-axis. And we
consider the plane of spins, oriented either in the $y-z$ plane or in the
$x-y$ plane. The results of computer simulation for $N=144$ spins are
presented in Fig. 7. These results demonstrate that the maximally
efficient spin reversal happens for the chain of spins directed along the
$x$-axis. This looks quite understandable, since the $x$-axis is the axis
of the direction of the resonator feedback field, which is the main source
of the coherent spin motion.

\section{Discussion}

We have accomplished computer simulations of the coherent spin relaxation
in molecular magnets with large spin. The investigation is based on the
microscopic model taking into account realistic dipole spin interactions
and the single-site magnetic anisotropy. The system is prepared in a
strongly nonequilibrium state, with an external magnetic field opposite to
the sample magnetization.

The principal point of our investigation is the presence of a resonator
coupled to the sample. The later is inserted into a coil of an electric
circuit, whose natural frequency is in resonance with the Zeeman frequency.
Without the resonator, the coherent spin motion is impossible. It is the
resonator feedback field, which collectivizes the spin motion, making it
well correlated, hence, coherent.

To realize the coherent spin relaxation, the Zeeman frequency $\om_0$ has
to be much larger than the anisotropy frequency $\om_D$. An efficient spin
reversal requires that the initial spin polarization be sufficiently high,
the higher, the better, The typical spin reversal time $\tau$ is an order
smaller than the transverse dephasing time $T_2$, which translates into
the relation $\gm\sim 10\gm_2$.

The role of dipole spin interactions, {\it in the presence of a resonator},
is twofold, making the spin dynamics in a sample coupled to a resonator
rather different from that happening in a sample with no resonator feedback
fields.

From one side, dipole interactions influence the spin
motion by making the spin reversal less pronounced. For low spins, such
as $S=1/2$, dipole interactions are less important than for large spins
$S=10$. Emphasizing the decoherence influence of the dipolar interactions,
we should keep in mind that our simulations are performed for a finite
number of spins. The majority of our calculations are done for $125$
spins. Because of the long range of the dipolar forces, increasing the
number of spins strengthens the decohering influence of these forces.
However, all qualitative results remain, as we have checked by varying
the number of spins between $64$ and $343$. Also, presenting the results
in dimensionless units, as we have done, when all frequencies and
attenuation parameters are normalized by the dipolar interaction strength,
makes the calculated curves for the average magnetization practically
independent of the sample size. Increasing the number of interacting spins
simply implies the renormalization of the dimensionless quantities and
does not change their behavior represented in dimensionless units.

At the same time, increasing the number $N$ of spins strengthens the role
of the resonator feedback field, which makes the process of spin
coherentization faster, so that the relaxation time, due to the coherent
spin motion, depends on the number of spins as $1/N$.

In this way, stronger dipole interactions, from one side, increase the
transverse decoherence attenuation, but, from another side, they induce
stronger coherence through the action of the resonator feedback field,
making the coherent relaxation faster. These two opposite effects, to some
extent, compensate each other. Therefore, the coherent spin dynamics,
occurring in the presence of a resonator, qualitatively does not change
much under the variation of spin number.

Our main concern in the present paper has been the study of spin dynamics
for large spins. This is why we have done numerical simulations for
$S=10$. We have had no aim of studying the low-spin dynamics, such as that
of spins one-half, since this dynamics has been considered earlier. It is
only to note that the low-spin dynamics is rather different that we show
its qualitative difference in one curve of Fig. 6.

It is worth mentioning that for large spins $S\geq 1$ it is feasible to
consider spin transitions between different sublevels labelled by the
$z$-projection number $m=-S,-S+1,\ldots,S-1,S$. A pair of sublevels can be
treated as an effective two-level system [56,57]. Then to realize the
coherent spin relaxation, one has to tune the resonant natural frequency
to the transition frequency between the chosen two levels. For high
nuclear spins, this procedure was realized experimentally [15,16]. Hence,
in the same way it can be realized for molecular magnets.

It is important to stress that there are several principle physical
differences between the experiments with nuclear spins, described in Refs.
[15,16], and the situation considered in the present paper. In these
experiments [15,16], the nuclei of $^{27}$Al inside the ruby crystal were
studied. {\it First} of all, the nuclei of  $^{27}$Al possess spins
$I=5/2$, which are not as large as we have considered here, dealing with
$S=10$. {\it Second}, in the case of $^{27}$Al, an external resonant
circuit was tuned to the central line $\{-1/2,1/2\}$, with a fixed
transition frequency $\om_0\sim 10^8$ Hz, thus, reducing the consideration
to an effective two-level system with spin one-half, while here we always
have dealt with the total spin $S=10$, since we have considered the
resonant circuit tuned to the transition between $-S$ and $S$. {\it Third},
as we have shown, for our high-spin case the influence of dipolar
interactions is essential, while their role for an effective one-half spin
system [15,16] is not of such importance. {\it Fourth}, contrary to the
case of nuclear spins, having no single-site anisotropy, the molecular
magnets, we have studied, exhibit quite strong magnetic anisotropy,
fundamentally distorting spin dynamics and making it principally different
as compared with the isotropic case of nuclear spins. {\it Fifth}, for
strongly polarized spin materials, it is necessary to take account of the
saturation effect making the total transverse attenuation depending on the
polarization level [1,25], as in Eq. (11), while this effect does not play
role for not so strong polarization [15,16]. {\it Sixth}, in experiments
[15,16], pulsing spin dynamics was analyzed, when the inversion of spin
polarization was permanently supported by constantly applied dynamic
nuclear polarization with a rather high pumping rate, while we studied the
pure coherent spin relaxation, when there is no permanent pumping.
{\it Finally}, we have studied here the dependence of spin relaxation on
the sample shape and orientation. Such geometric effects, to our knowledge,
have not been investigated. These seven factors make the spin dynamics in
our case and in the case of Refs. [15,16] principally different.

When studying the geometric effects related to the sample shape
and its orientation, differently oriented spin chains and planes have been
considered. We have found that, under the same system parameters, including
the number of spins, except the sample shape and orientation, the most
efficient coherent spin relaxation, with the deepest spin reversal, happens
for a chain of spins aligned with the direction of the resonator feedback
field.

The main aim of the present paper has been to analyze the coherent spin
relaxation under widely varying system parameters, in order to clarify the
influence of different parameters on the coherent spin motion. This should
help in choosing the optimal materials for realizing such a coherent spin
motion. Nowadays, there are plenty of magnetic materials, with widely
varying properties, which could be used for experimentally observing the
described effects. The description of the properties of various molecular
magnets can be found in the review articles [5,18,44,45].

As an example, we can mention the most often studied molecular magnets
made of the molecules Mn$_{12}$ or Fe$_8$, whose spins are $S=10$. For
these materials, $\om_D\sim 10^{12}$ s$^{-1}$. At low temperatures, below
the blocking temperature of about 1 K, the sample can be polarized, having
a very long spin-lattice relaxation time $T_1\sim 10^5 - 10^7$ s. Hence
$\gm_1\equiv 1/T_1$ is practically negligible, $\gm_1\sim 10^{-7}-10^{-5}$
s$^{-1}$. Dipole spin interactions in these materials are rather strong,
with $\gm_2\sim 10^{10}$ s$^{-1}$. To realize the coherent spin relaxation,
the external magnetic field $B_0$ is to be sufficiently strong, such that
corresponding Zeeman frequency $\om_0$, being close to the resonator
natural frequency $\om$, would be much larger than the anisotropy
frequency $\om_D$. For the considered case of Mn$_{12}$ or Fe$_8$, this
requires the field $B_0\sim 100$ T. This is a strong field, though which
can experimentally be reached [60]. Fortunately, there are many other
molecular magnets with a smaller anisotropy. For instance, in the case of
nanomagnets formed by the molecules Mn$_6$, whose spin is $S=12$, the
magnetic anisotropy is much lower, with $\om_D\sim 10^{10}$ s$^{-1}$.
Therefore the required external magnetic field is only about $B_0\sim 1$
T, which is the standard field used in laboratory.

The existence of the magnetic anisotropy typical of many molecular magnets,
hinders the feasibility of the coherent spin relaxation. However, there
are many magnetic materials with a small anisotropy, which should not
disturb the coherent spin motion. In addition, the influence of the
magnetic anisotropy can always be suppressed by a sufficiently strong
external magnetic field. Fortunately, there exists in the world the
possibility of creating very strong magnetic fields. Among available
sources [60], we may mention those where the magnetic fields up to $45\;T$
(USA) and even $600\; T$ (Japan) can be reached.

In order to estimate the typical time of the coherent spin relaxation, we
may notice that this time is an order smaller than $T_2$. The spin
dephasing time in molecular magnets, such as Mn$_{12}$ and Fe$_8$, is due
to dipole interactions yielding $\gm_2\sim 10^{10}$ s$^{-1}$. Hence
$T_2\sim 10^{-10}$ s. This means that the typical time of the coherent
spin relaxation in these materials is $10^{-11}$ s, which is an ultrafast
process.

\vskip 2mm

In conclusion, it is worthwhile to mention that for many magnetic
molecules the influence of hyperfine interactions from nuclear spins
could be important. These interactions result in the appearance of an
additional line broadening, which can be included in the effective
attenuation parameters, so that the existence of the hyperfine
interactions can be taken into account by the appropriate definition
of the effective attenuations, as has been done in Ref. [42]. At the
same time, for many molecules, typical of the large family of magnetic
molecules, such as Mn$_{12}$ and Fe$_8$, the hyperfine interactions are
of the order of $10^{-3}$ K, which are much weaker than the dipolar
interactions, being of the order of $0.1$ K (see details in the review
articles [5,18,44,45]). When the hyperfine interactions are two orders
smaller than the dipolar interactions, the former, with a very good
approximation, can be neglected. And if the former are comparable with
the latter, this can be taken into account by the corresponding definition
of the line broadening.

\vskip 5mm

{\bf Acknowledgement}

\vskip 2mm

One of the authors (V.I.Y.) is grateful for useful discussions to B.
Barbara and E. Yukalova. And the other authors (V.K.H. and P.V.K.)
appreciate fruitful discussions with Y.L. Raikher. Financial support
under the grant of RFBR 07-02-96026 is acknowledged.

\newpage

\begin{center}

{\bf{\Large Figure Captions} }
\end{center}

{\bf Fig. 1}. Reduced spin variable $s$, for a cubic lattice,
characterizing the spin polarization along the $z$-axis, as a function of
dimensionless time (measured in units of $T_2$) for the Zeeman frequencies
$\om_0=1000$ (solid line), $\om_0=2000$ (long-dashed line), and
$\om_0=5000$ (short-dashed line). The simulation is done for the molecules
of spin $S=10$, with the reduced initial polarization $s_0=0.9$. The
anisotropy frequency is $\om_D=20$ and the resonator damping is $\gm=10$.

\vskip 5mm

{\bf Fig. 2}. Reduced spin polarization $s$, as a function of
dimensionless time, for a cubic lattice, with $\om_0=2000$ and
$\om_D=20$, for the resonator damping $\gm=1$ (solid line), $\gm=10$
(long-dashed line), and $\gm=50$ (short-dashed line). The sample of
molecules with spin $S=10$ has the initial polarization $s_0=0.9$.

\vskip 5mm

{\bf Fig. 3} Reduced spin polarization $s$, as a function of
dimensionless time, for a cubic lattice, with $\om_0=2000$,
$\om_D=20$, $\gm=10$, $S=10$, for different initial polarizations
$s_0=0.9$ (solid line), $s_0=0.7$ (long-dashed line), and $s_0=0.5$
(short-dashed line).

\vskip 5mm

{\bf Fig. 4} Spin polarization $s$, as a function of
dimensionless time, for a cubic lattice, with $\om_0=2000$,
$\gm=10$, $S=10$, and for different magnetic anisotropy values
characterized by the anisotropy frequency $\om_D=20$ (solid line),
$\om_D=50$ (long-dashed line), $\om_D=100$ (short-dashed line).

\vskip 5mm

{\bf Fig. 5}. Spin polarization $s$, as a function of
dimensionless time, for a cubic lattice, with $\om_0=2000$,
$\om_D=20$, $\gm=10$, and $S=10$, for two different cases, when the
dipole interactions are present (solid line) and when they are absent
(dashed line).

\vskip 5mm

{\bf Fig. 6}. Spin polarization $s$, as a function of
dimensionless time, for a cubic lattice, with $\om_0=2000$,
$\om_D=20$, $\gm=1$, and for different spins $S=10$ (solid line)
and $S=1/2$ (dashed line).

\vskip 5mm

{\bf Fig. 7}. Difference in the behavior of spin relaxation for different
sample shapes and orientations, under the same values $\om_0=2000$,
$\om_D=20$, $\gm=30$, $S=10$. The chain of spins along the $z$-axis (solid
line); the chain of spins along the $x$-axis (long-dashed line); the plain
of spins in the $y-z$ plane (short-dashed line), and the plane of spins in
the $x-y$ plane (dashed-dotted line).

\newpage

\begin{figure}[ht]
\centerline{\psfig{file=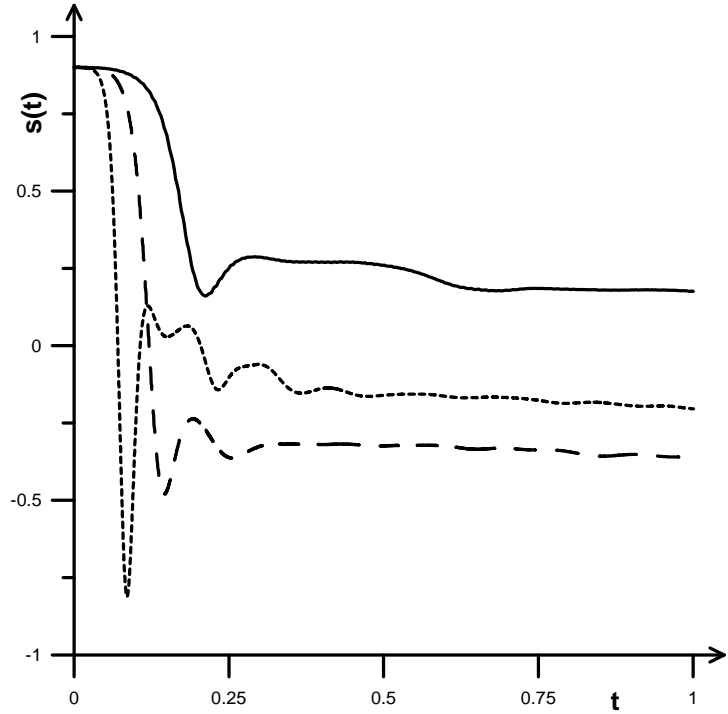,angle=0,width=10cm}}
\caption{Reduced spin variable $s$, for a cubic lattice,
characterizing the spin polarization along the $z$-axis,
as a function of dimensionless time (measured in units of $T_2$)
for the Zeeman frequencies $\om_0=1000$ (solid line), $\om_0=2000$
(long-dashed line), and $\om_0=5000$ (short-dashed line). The
simulation is done for the molecules of spin $S=10$, with the
reduced initial polarization $s_0=0.9$. The anisotropy frequency
is $\om_D=20$ and the resonator damping is $\gm=10$.
}
\label{fig:Fig.1}
\end{figure}

\begin{figure}[ht]
\centerline{\psfig{file=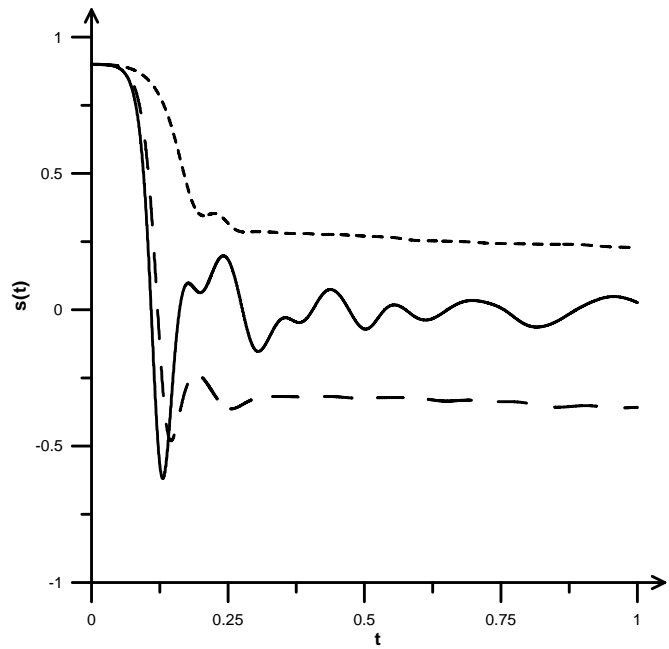,angle=0,width=10cm}}
\caption{ Reduced spin polarization $s$, as a function of
dimensionless time, for a cubic lattice, with $\om_0=2000$ and
$\om_D=20$, for the resonator damping $\gm=1$ (solid line), $\gm=10$
(long-dashed line), and $\gm=50$ (short-dashed line). The sample of
molecules with spin $S=10$ has the initial polarization $s_0=0.9$.
}
\label{fig:Fig.2}
\end{figure}

\newpage

\begin{figure}[ht]
\centerline{\psfig{file=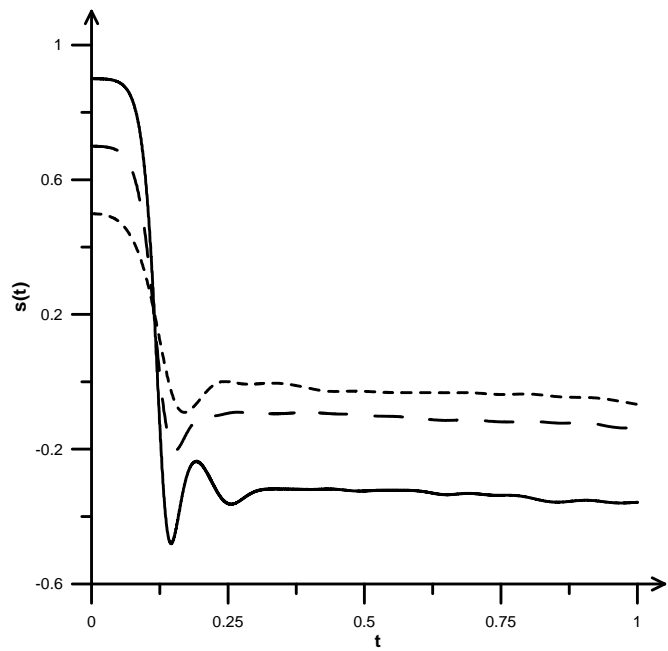,angle=0,width=10cm}}
\caption{ Reduced spin polarization $s$, as a function of
dimensionless time, for a cubic lattice, with $\om_0=2000$,
$\om_D=20$, $\gm=10$, $S=10$, for different initial polarizations
$s_0=0.9$ (solid line), $s_0=0.7$ (long-dashed line), and $s_0=0.5$
(short-dashed line).
}
\label{fig:Fig.3}
\end{figure}

\begin{figure}[ht]
\centerline{\psfig{file=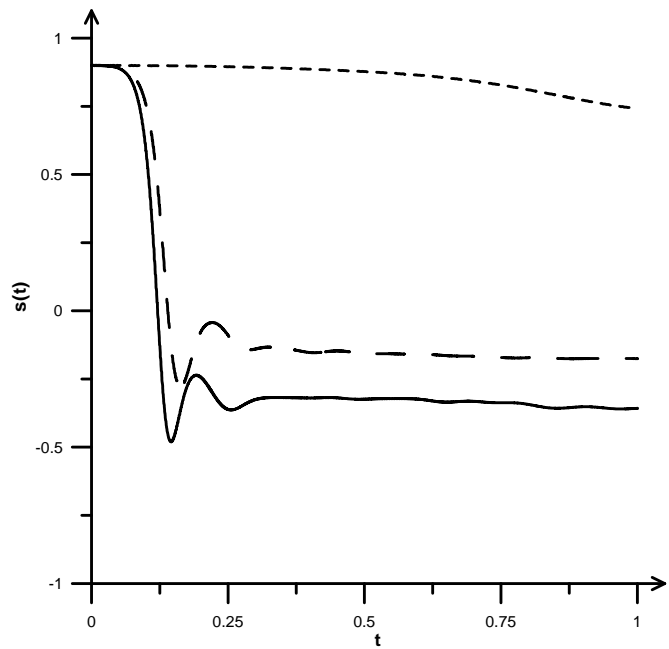,angle=0,width=10cm}}
\caption{Spin polarization $s$, as a function of
dimensionless time, for a cubic lattice, with $\om_0=2000$,
$\gm=10$, $S=10$, and for different magnetic anisotropy values
characterized by the anisotropy frequency $\om_D=20$ (solid line),
$\om_D=50$ (long-dashed line), $\om_D=100$ (short-dashed line).
}
\label{fig:Fig.4}
\end{figure}

\newpage

\begin{figure}[ht]
\centerline{\psfig{file=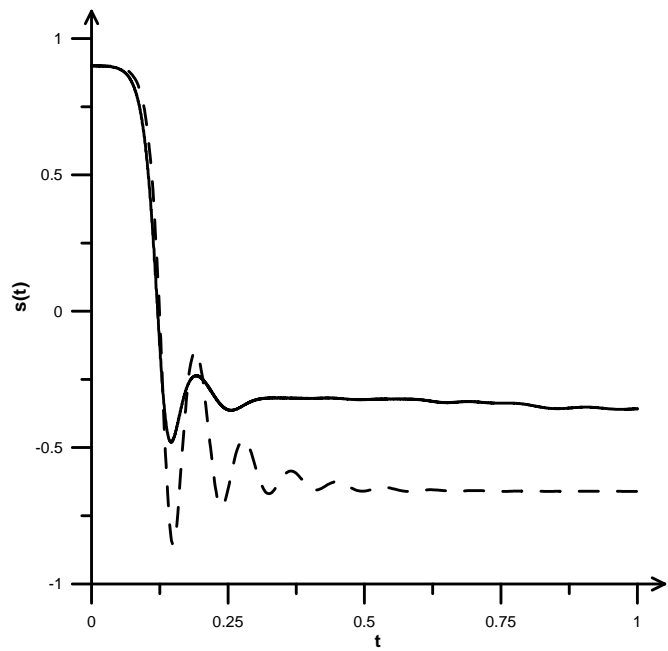,angle=0,width=10cm}}
\caption{Spin polarization $s$, as a function of
dimensionless time, for a cubic lattice, with $\om_0=2000$,
$\om_D=20$, $\gm=10$, and $S=10$, for two different cases, when the
dipole interactions are present (solid line) and when they are absent
(dashed line).
}
\label{fig:Fig.5}
\end{figure}

\begin{figure}[ht]
\centerline{\psfig{file=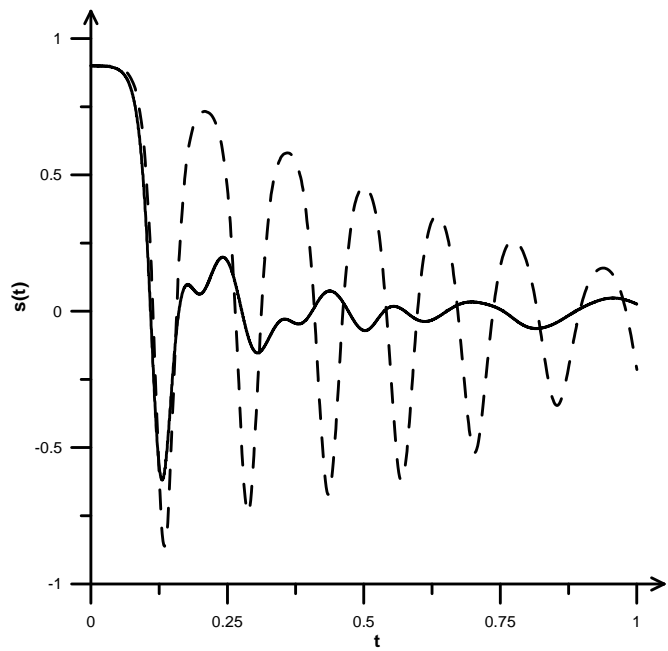,angle=0,width=10cm}}
\caption{Spin polarization $s$, as a function of
dimensionless time, for a cubic lattice, with $\om_0=2000$,
$\om_D=20$, $\gm=1$, and for different spins $S=10$ (solid line)
and $S=1/2$ (dashed line).
}
\label{fig:Fig.6}
\end{figure}

\newpage

\begin{figure}[ht]
\centerline{\psfig{file=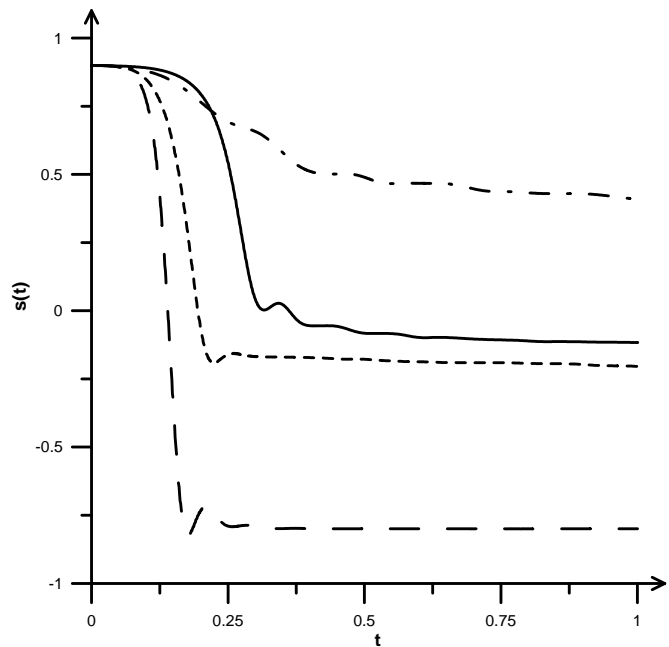,angle=0,width=10cm}}
\caption{Difference in the behavior of spin relaxation for different
sample shapes and orientations, under the same values $\om_0=2000$,
$\om_D=20$, $\gm=30$, $S=10$. The chain of spins along the $z$-axis (solid
line); the chain of spins along the $x$-axis (long-dashed line); the plain
of spins in the $y-z$ plane (short-dashed line), and the plane of spins in
the $x-y$ plane (dashed-dotted line).
}
\label{fig:Fig.7}
\end{figure}

\end{document}